\documentclass[aps,pra,reprint,groupedaddress,twocolumn,floatfix]{revtex4-2}

\usepackage{graphicx}
\graphicspath{ {./images/} }
\usepackage{xcolor}
\usepackage{amsmath}
\usepackage{mathtools}
\usepackage{amssymb}
\usepackage{flushend}
\usepackage[T1]{fontenc}
\usepackage{placeins}

\usepackage{braket}
\usepackage{amssymb}
\usepackage{amsmath}
\usepackage{natbib}
\usepackage{braket}
\usepackage{amssymb}
\usepackage{amsmath}
\usepackage{dsfont}
\usepackage{algorithm}
\usepackage{algpseudocode}
\makeatletter
\def\ftype@algorithm{8}
\expandafter\newbox\csname fbox@\ftype@algorithm\endcsname
\expandafter\setbox\csname fbox@\ftype@algorithm\endcsname\hbox{}
\makeatother
\usepackage{float}

\begin{document}

\title{Improved lower bound for the two-way-assisted \\ quantum capacity of the bosonic thermal-loss channel}

\author{Matthew Barber}
\affiliation{Department of Computer Science, University of York, York YO10 5GH, United Kingdom}
\author{Stefano Pirandola}
\affiliation{Department of Computer Science, University of York, York YO10 5GH, United Kingdom}

\begin{abstract}
The bosonic thermal-loss channel is a fundamental model for quantum communication. This Gaussian channel models the transmission of bosonic systems subject to both loss and thermal noise.
It describes many practical systems, including optical fibres, waveguides and free-space links, where background thermal noise is significant. Determining the channel's two-way-assisted quantum capacity, the maximum rate at which quantum information can be transmitted reliably through the channel with two-way classical assistance, remains an open problem. Here, we improve upon the best known lower bound for this capacity for a wide range of channel parameters. We achieve this by combining an improved qubit-bosonic distribution technique with a recently introduced technique for discovering entanglement distillation protocols.
\end{abstract}

\maketitle

\section{Introduction}

The classical internet has transformed the world by allowing distant computers to communicate efficiently.
However, it is restricted to classical data consisting of discrete bits, each taking the value $0$ or $1$~\cite{CoverThomas2006,mackay2003information,bertsekas1992data,elgamal2011network}.
Quantum physics provides a much richer framework for encoding information in physical systems~\cite{nielsen2000quantum,Simeone_2026,Serafini2023}.
A quantum internet~\cite{Kimble2008QuantumInternet,PirandolaBraunstein2016,QuantumInternet}, a network capable of transmitting quantum information, could supplement classical networks, enabling new applications such as provably secure private communication~\cite{QuantumCryptography} and distributed quantum computing~\cite{Barral2025DistributedQuantumComputing}.
\par
The preferred carrier of quantum information is the photon.
Photons can be transmitted optically through fibres and free-space links.
At lower frequencies, such as the terahertz (THz) and microwave regimes, they can be transmitted via appropriate waveguides and wireless communication links.
Under suitable assumptions, each of these scenarios reduces to a thermal-loss channel $\mathcal{E}_{\eta, \bar{n}}$, a bosonic Gaussian channel characterized by two parameters: its transmissivity $\eta \in \left[0, 1\right]$ and its thermal number $\bar{n} \in \left[0, \infty\right)$.
The transmissivity represents the fraction of input photons that reach the output.
The thermal number represents the mean number of photons present in the environment, a fraction $1-\eta$ of which enters the channel.
\par
In the optical range, the environmental thermal bath is negligible.
However, fibre imperfections introduce excess noise $\xi$, which maps to an equivalent thermal number $\bar{n}$ via $\xi=\eta^{-1} (1-\eta) \bar{n}$~\cite[Eq.~(109)]{LB2}.
For free-space optical links with line-of-sight transmissivity $\eta$, $\bar{n}$ characterizes background skylight collected by the receiver, assuming negligible atmospheric turbulence and pointing errors~\cite{PhysRevResearch.3.013279}.
At THz and microwave frequencies, the natural thermal background is significant; $\bar{n}$ increases with wavelength, although spatio-temporal filters partially mitigate this~\cite{PhysRevResearch.3.043014}. 
\par
Given the widespread use of this model, a fundamental question is how much quantum information can be transmitted per use of a thermal-loss channel $\mathcal{E}_{\eta, \bar{n}}$ when the communicating parties are assisted by local operations and two-way classical communication (CC).
The corresponding information-theoretic quantity is the two-way-assisted (or simply `two-way') quantum capacity, $Q_2(\mathcal{E}_{\eta, \bar{n}})$.
Furthermore, under two-way CC, sending a qubit and distributing an entanglement bit (or `ebit') are equivalent operations, so $Q_2(\mathcal{E}_{\eta, \bar{n}})=D_2(\mathcal{E}_{\eta, \bar{n}})$, where $D_2$ is the two-way entanglement distribution capacity.
In turn, $D_2$ lower-bounds the two-way secret-key capacity $K_2$ of the channel, because distributing an ebit enables the generation of a shared secret bit.
\par
Determining $Q_2(\mathcal{E}_{\eta, \bar{n}})$ in general remains an open problem, though both upper and lower bounds exist.
The best known upper bound is~\cite{PLOB}
\begin{equation}
Q_2(\mathcal{E}_{\eta,\bar{n}})
\le
\begin{cases}
-g(\bar{n})
-\log_2\!\left[(1-\eta)\eta^{\bar{n}}\right],
& \displaystyle \bar{n}<\frac{\eta}{1-\eta},
\\[1.2ex]
0,
& \displaystyle \bar{n}\geq\frac{\eta}{1-\eta},
\end{cases}\label{PLOBUpperBound}
\end{equation}
where, for every $x \in \left(0, \infty\right)$,
\begin{equation}
g\left(x\right) := \left(x+1\right)\log_{2}\left(x+1\right)-x\log_{2}\left(x\right).
\end{equation}
Likewise, Ref.~\cite{ReverseCoherentInformation} extended the reverse coherent information (RCI) to bosonic systems to set the lower bound
\begin{equation}
Q_2\left(\mathcal{E}_{\eta, \bar{n}}\right) \geq \max\left\{0, - g\left(\bar{n}\right) - \log_{2}\left(1-\eta\right)\right\}.\label{RCILowerBound}
\end{equation}
Note that, if  $-g\left(\bar{n}\right) - \log_{2}\left(1-\eta\right) > 0$ and $-\bar{n}\log_{2}\eta \ll -g\left(\bar{n}\right) - \log_{2}\left(1-\eta\right)$,
the bounds in Eqs.~\eqref{PLOBUpperBound} and~\eqref{RCILowerBound} nearly coincide.
However, outside this regime, there will be quite a large gap between them.

To improve the lower bound in Eq.~\eqref{RCILowerBound}, Ref.~\cite{LowerBound} introduced a protocol for distributing entanglement over thermal-loss channels consisting of three steps.
First, Alice produces a hybrid qubit-bosonic quantum state and sends half of it through the channel.
Second, upon receiving his half, Bob performs a projective measurement.
These initial steps depend on two parameters which are optimized to maximize the overall rate of the protocol.
Finally, if Bob's measurement results in a particular outcome, Alice and Bob perform an entanglement distillation protocol.
The rate of the overall scheme then provides a lower bound on $D_2(\mathcal{E}_{\eta, \bar{n}})$ and thus $Q_2(\mathcal{E}_{\eta, \bar{n}})$.
Across much of the regime in which the bounds in Eqs.~\eqref{PLOBUpperBound} and~\eqref{RCILowerBound} differ substantially, this lower bound improves upon the bound in Eq.~\eqref{RCILowerBound}~\footnote{Note that Eq.~\eqref{PLOBUpperBound} from Ref.~\cite{PLOB} shows that $\eta > \frac{\bar{n}}{\bar{n}+1}$ is a necessary condition for $Q_{2}>0$. Ref.~\cite{LowerBound} showed that it is also a sufficient condition. Combining these two conditions, one has that the two-way quantum capacity of the thermal-loss channel is positive if and only if $\eta > \frac{\bar{n}}{\bar{n}+1}$.}.

\par
Here, we introduce an improved entanglement distribution protocol able to reduce the gap between lower and upper bounds. First, we generalize the initial two steps of the protocol in Ref.~\cite{LowerBound}, introducing a total of four parameters.
We then replace the distillation protocol used in Ref.~\cite{LowerBound} with the heuristic lookahead distillation method from Ref.~\cite{HeuristicSearch}.
This results in a significantly improved lower bound on the two-way quantum capacity of the thermal-loss channel, outperforming the main lower bound of Ref.~\cite{LowerBound} across the full range of channel parameters tested.
Our results reduce the gap between the lower and upper bounds for both $Q_2(\mathcal{E}_{\eta, \bar{n}})$ and $K_2(\mathcal{E}_{\eta, \bar{n}})$~\cite{LB1,LB2,LB3,LB4}.
Furthermore, we demonstrate similar improvements for the two-way capacities of other bosonic Gaussian channels key to continuous-variable quantum information theory, specifically the quantum amplifier and the Gaussian additive-noise channel.

\section{The entanglement distribution protocols}\label{Protocol}

\subsection{Basic distribution protocol}

Ref.~\cite{LowerBound} proposed the following protocol to distribute entanglement across thermal-loss channels.
For some $M \in \mathbb{N}$ and $c \in \left(0, 1\right)$, Alice prepares the state
\begin{equation}
    \Ket{\Phi_{M, c}} = c\Ket{e_{0}}\Ket{0} + \sqrt{1-c^{2}}\Ket{e_{1}}\Ket{M},
\end{equation}
where $\Ket{e_{0}}$ and $\Ket{e_{1}}$ are orthogonal qubit states in her register, while $\Ket{0}$ and $\Ket{M}$ are bosonic number states with $0$ and $M$ photons, respectively.
The bosonic half of Alice's system is then transmitted to Bob via $\mathcal{E}_{\eta, \bar{n}}$.
Upon receipt, Bob performs a measurement described by the positive operator-valued measure (POVM) $\left\{\Pi_{M}, \mathds{1} - \Pi_{M}\right\}$, where $\mathds{1}$ is the identity operator on the bosonic system and $\Pi_{M} = \Ket{0}\Bra{0} + \Ket{M}\Bra{M}$.
If the bosonic system falls outside the support of $\Pi_{M}$, it is discarded.
Otherwise, $\Ket{0}$ is mapped to $\Ket{f_{0}}$ and $\Ket{M}$ to $\Ket{f_{1}}$, where $\Ket{f_{0}}$ and $\Ket{f_{1}}$ are orthogonal qubit states in Bob's register.
Alice and Bob now share a qubit pair, to which they apply a Pauli-based twirling operation~\cite{Bennett1996Mixed} that preserves its diagonal elements in the Bell basis while eliminating the off-diagonal terms.
They then apply entanglement distillation~\cite{PhysRevA.98.042309,PhysRevA.71.062325}, converting their Bell-diagonal states into perfect ebits.
In this scheme, $c$ and $M$ are optimized to maximize the overall entanglement yield.

\subsection{Improved qubit-bosonic distribution}\label{improvedQB}

Our distribution protocol involves a continuous parameter $c \in \left(0, 1\right)$ and three integer parameters $\left(M_{1}, M_{2}, N\right) \in \left\{0, 1, 2, \dots\right\}^{3}$ with 
\begin{equation}
    M_{1} < M_{2}.
\end{equation}
Alice prepares the state $\Phi:=\Ket{\Phi_{M_{1}, M_{2}, c}}\Bra{\Phi_{M_{1}, M_{2}, c}}$, where
\begin{equation}
    \Ket{\Phi_{M_{1}, M_{2}, c}} = c\Ket{e_{0}}\Ket{M_{1}} + \sqrt{1-c^{2}}\Ket{e_{1}}\Ket{M_{2}},
\end{equation}
and sends the bosonic part through the thermal-loss channel $\mathcal{E}_{\eta, \bar{n}}$.
At the channel output, Bob measures the bosonic mode with the POVM $\left\{\Pi, \mathds{1} - \Pi \right\}$, where $\Pi$ is
\begin{equation}
   \sum_{n=\max\left(\left\{0, N+1-M\right\}\right)}^{N}\Ket{n}\Bra{n} + \sum_{n=\max\left(\left\{M, N+1\right\}\right)}^{N+M}\Ket{n}\Bra{n}
\end{equation}
and $M = M_{2}-M_{1}$.
The output is retained only if projected onto the support of $\Pi$.
\par
The conditional state is then transformed into a qubit pair via the following steps.
First, Bob applies a continuous-to-discrete variable transformation.
For each $m \in \left\{\max\left(\left\{0, N+1-M\right\}\right), \dots, N\right\}$, he maps $\Ket{m}$ to $\Ket{f_{0}}\Ket{g_{m}}$ and, for $m \in \left\{\max\left(\left\{M, N+1\right\}\right), \dots, N+M\right\}$, he maps $\Ket{m}$ to $\Ket{f_{1}}\Ket{g_{m-M}}$, where
\begin{equation}
    \left\{\Ket{g_{\max\left(\left\{0, N+1-M\right\}\right)}}, \dots, \Ket{g_{N}}\right\}\label{auxiliaryBasis}
\end{equation}
are orthogonal states of an auxiliary discrete-variable register belonging to Bob.
Second, he measures his auxiliary register in the basis of Eq.~\eqref{auxiliaryBasis} and sends the result to Alice.
For every $m \in \left\{\max\left(\left\{0, N+1-M\right\}\right), \dots, N\right\}$, let the probability of the bosonic mode being found to lie in the support of $\Pi$ and the auxiliary register being measured as $\Ket{g_{m}}$ be $p_{m}$ and the distributed state conditional on these measurement outcomes be $\rho^{\left(m\right)}$.
For each such $m$, this distributed state can be expressed in the Bell basis $\left\{\Ket{\Phi_{0,0}}, \Ket{\Phi_{0,1}}, \Ket{\Phi_{1,0}}, \Ket{\Phi_{1,1}}\right\}$ as
\begin{equation}
    \rho^{\left(m\right)} = \sum_{\left(i, j, k, l\right) \in \left\{0, 1\right\}^{4}}\rho_{i, j, k, l}^{\left(m\right)}\Ket{\Phi_{i, j}}\Bra{\Phi_{k, l}},
\end{equation}
where the elements of the density matrix $\rho_{i, j, k, l}^{\left(m\right)}$ are given in Appendix~\ref{OutputComputation}. 

\begin{figure*}
    \centering
    \includegraphics[width=1\textwidth]{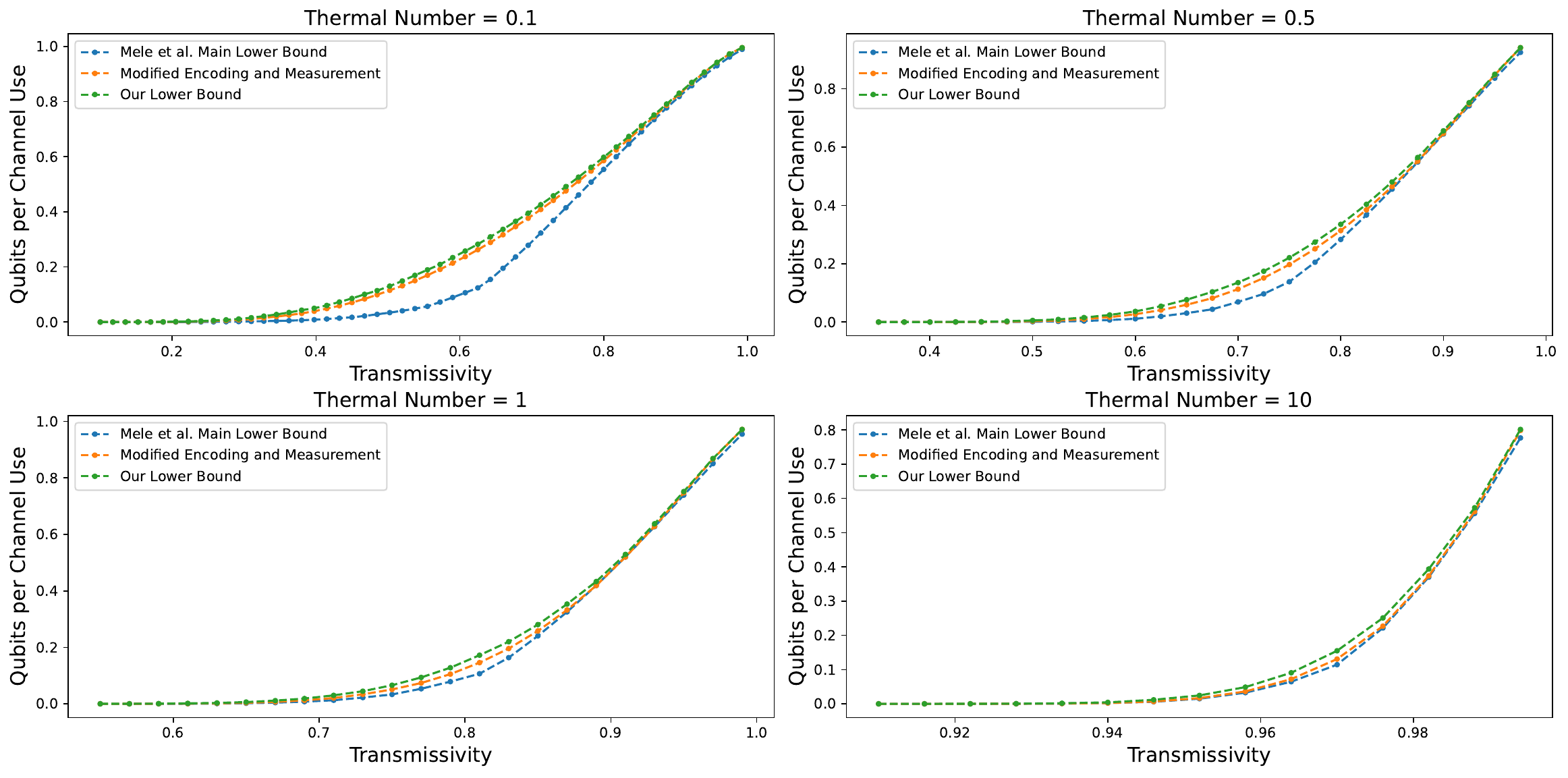}
    \caption{Lower bounds for the two-way quantum capacity $Q_2$ of the thermal-loss channel $\mathcal{E}_{\eta,\bar{n}}$. We plot the achievable rate (in qubits per channel use) versus the transmissivity $\eta$ at various thermal numbers, from $\bar{n}=0.1$ to $10$. Our lower bound (green) is compared with the lower bound of Ref.~\cite{LowerBound} described in Section~\ref{Protocol} (blue). 
    For reference, we also show the performance of our distribution protocol when employing our improved encoding and measurement method with the distillation protocol used in Ref.~\cite{LowerBound}, rather than that of Ref.~\cite{HeuristicSearch} (orange). Dashed lines show linear interpolations between neighbouring points; these are only intended as visual guides, not bounds themselves.}
    \label{fig:LowerBoundOnly}
\end{figure*}

\begin{figure*}
    \centering
    \includegraphics[width=1\textwidth]{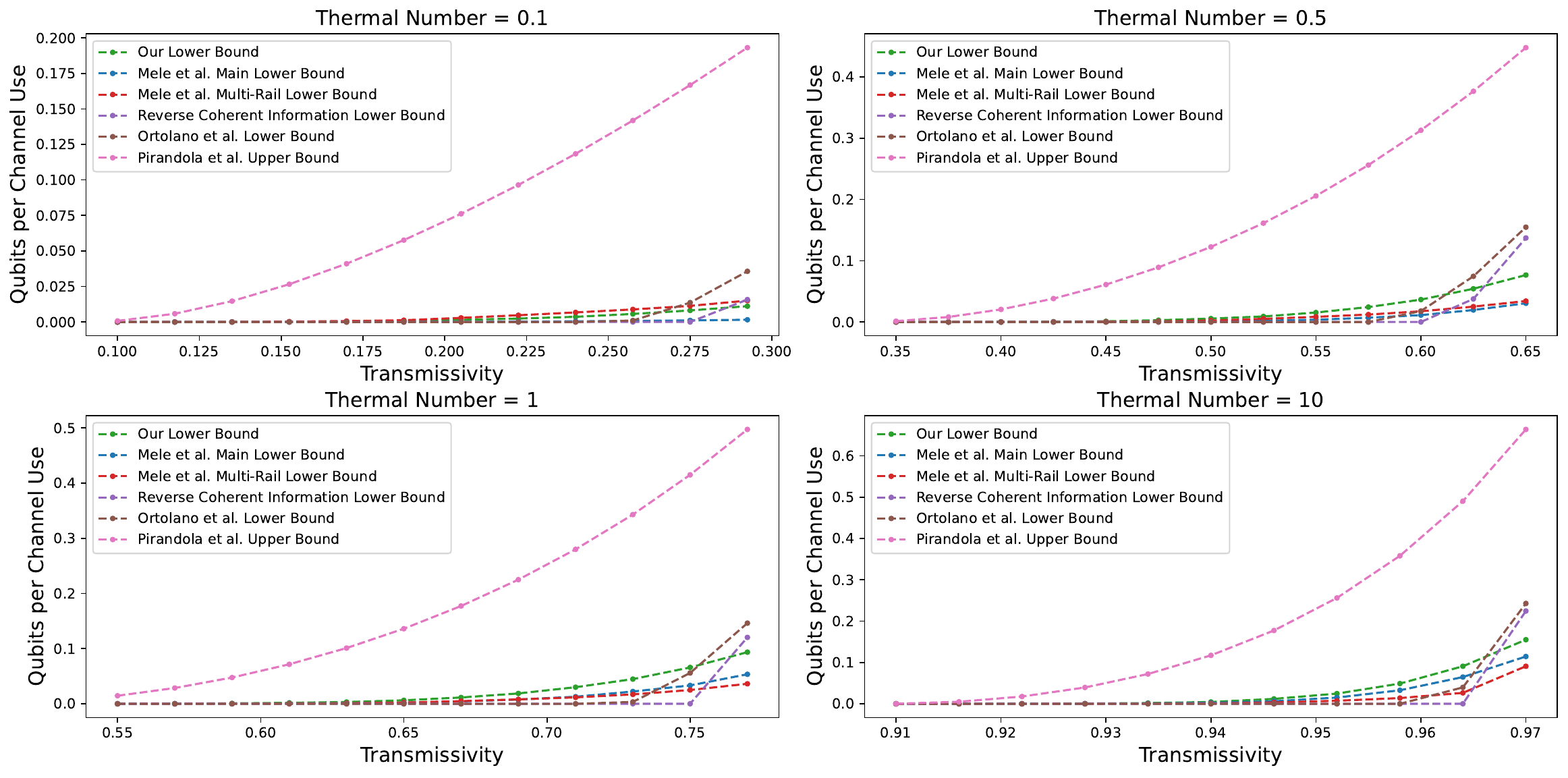}
    \caption{Bounds on the two-way quantum capacity of the thermal-loss channel $\mathcal{E}_{\eta,\bar{n}}$. We plot these bounds versus the transmissivity $\eta$ at various thermal numbers, from $\bar{n}=0.1$ to $10$. Our lower bound (green) is compared with the main lower bound from Ref.~\cite{LowerBound}, described in Section~\ref{Protocol} (blue), the RCI lower bound of Ref.~\cite{ReverseCoherentInformation} (purple) and the PLOB thermal upper bound of Ref.~\cite{PLOB} (pink). For completeness, we also plot the multi-rail lower bound of Ref.~\cite{LowerBound} (red) and the secret-key capacity lower bound of Ref.~\cite{LB4} (brown), which does not apply to $Q_2$ but $K_2$. 
    Dashed lines show linear interpolations between points; these are only intended as visual guides, not bounds themselves. 
    }
    \label{fig:FullPictureRestricted}
\end{figure*}

\subsection{Improved distillation of the output pairs}

Suppose Alice and Bob perform the improved distribution protocol $N \gg 1$ times.
Asymptotically, for every $m \in \left\{\max\left(\left\{0, N+1-M\right\}\right), \dots, N\right\}$, they collect $Np_{m}$ output pairs in the state $\rho^{\left(m\right)}$, which are subjected to an entanglement distillation protocol with yield $Y\left(\rho^{\left(m\right)}\right)$.
The entanglement distribution rate is therefore
\begin{equation}
    \sum_{m=\max\left(\left\{0, N+1-M\right\}\right)}^{N}p_{m}Y\left(\rho^{\left(m\right)}\right)\text{,}
\end{equation}
which serves as a lower bound on $D_2\left(\mathcal{E}_{\eta, \bar{n}}\right) = Q_2\left(\mathcal{E}_{\eta, \bar{n}}\right)$.
\par
In the prior work of Ref.~\cite{LowerBound}, the authors employed four steps to distil their pairs.
First, they applied a Pauli-based twirling operation to convert their distributed qubit pairs into Bell-diagonal states while preserving their diagonal components in the Bell basis.
Second, they applied the P1-or-P2 recurrence protocol of Ref.~\cite{PhysRevA.98.042309} an optimal number of times.
Third, they performed an optimal permutation of the four Bell states.
Fourth, they applied the 2-copy protocol of Ref.~\cite{PhysRevA.71.062325}.
In the second and third steps, optimality meant adopting the choice that maximized the overall yield.
\par
In our protocol, we retain the initial twirling step.
Then, rather than using the recurrence protocol of Ref.~\cite{PhysRevA.98.042309}, we apply an optimal number of iterations of the protocol from Ref.~\cite{Recurrence}.
Before each application of this recurrence protocol, we permute the four Bell states to maximize the Bell fidelity of the output pairs upon success.
Finally, we apply the $(n, r, d)$-lookahead distillation protocol of Ref.~\cite{HeuristicSearch} with $n = 4$, $r = 2$ and $d = 3$, in the terminology of that paper.
\par
The protocol of Ref.~\cite{HeuristicSearch} distils entanglement from blocks of $n$ noisy Bell pairs by adaptively selecting parity checks implemented via appended-ebit measurements (AEMs) or bilateral Pauli measurements (BPMs).
At each step, the protocol searches over possible measurement sequences to a lookahead depth $d$ and performs the measurement expected to maximize the overall yield.
After each outcome, the state distribution and the sets of candidate parity checks are updated, and the search is repeated.
If only one undistilled pair remains, $r$ independent copies of this residual state are grouped into a new block before continuing.
Thus, $n$ controls the initial block size, $d$ controls the search depth and $r$ controls the regrouping of one-pair residual states.

\section{Improved lower bound}

In Fig.~\ref{fig:LowerBoundOnly}, we study the performance of our entanglement distribution protocol, comparing it with the main bound in Ref.~\cite{LowerBound}.
We obtain an improvement across the full range of parameters tested.
This stems largely from the modified encoding and measurement used in the qubit-bosonic distribution, as described in Section~\ref{improvedQB}, with additional performance gains achieved via the lookahead distillation protocol of Ref.~\cite{HeuristicSearch}.
The advantage of the lookahead protocol is evident at high thermal numbers.
As seen across the panels, the gain due to the modified encoding and measurement (orange) progressively collapses towards the bound of Ref.~\cite{LowerBound} (blue) as the thermal number increases.
However, when this modified approach is combined with lookahead distillation (green), the improvement over Ref.~\cite{LowerBound} remains stable.  

It is important to stress that the modified encoding and measurement scheme, parametrized by our four parameters, was optimized assuming the same distillation protocol as in Ref.~\cite{LowerBound}.
The resulting optimal parameters were then held fixed when the distillation protocol was replaced by the lookahead strategy of Ref.~\cite{HeuristicSearch}.
A full reoptimization of the four parameters specifically for the lookahead protocol is expected to yield even better performance, but is computationally demanding.

\par
In Fig.~\ref{fig:FullPictureRestricted}, we also compare the performance of our protocol with the RCI lower bound~\cite{ReverseCoherentInformation} of Eq.~\eqref{RCILowerBound} and the PLOB thermal upper bound~\cite{PLOB} of Eq.~\eqref{PLOBUpperBound}.
As shown in the figure, we outperform the RCI lower bound across a wider range of channel parameters than the previous bound of Ref.~\cite{LowerBound}.
Nevertheless, a substantial gap remains between these lower bounds and the PLOB thermal upper bound.
Closing the gap between the upper and lower bounds remains an active area of investigation. 
\par
For completeness, Fig.~\ref{fig:FullPictureRestricted} also includes the second, multi-rail bound proposed in Ref.~\cite{LowerBound}.
This bound performs well at low thermal numbers, such as $\bar{n}=0.1$, where it exceeds the main bound of Ref.~\cite{LowerBound} and slightly outperforms ours.
However, the multi-rail bound rapidly becomes inefficient as we move into the interesting regime with higher thermal numbers.
Finally, we have included the lower bound for the two-way secret-key capacity $K_2$ established in Ref.~\cite{LB4}.
To our knowledge, this is the best lower bound on $K_{2}$ for thermal-loss channels.
Because $K_2 \ge D_2 = Q_2$, this secret-key lower bound does not apply to the two-way quantum capacity.
However, our new lower bound can also be applied to $K_2$, providing improved performance at lower transmissivities.
\par
As mentioned earlier in our introduction, it is worth noting that our approach also improves the known lower bounds for the two-way capacities of other bosonic Gaussian channels.
In particular, we achieve similar improvements for the quantum amplifier and the Gaussian additive-noise channel, as discussed in Appendix~\ref{app:gchannel}.
Appendix~\ref{app:numerics} provides pseudocode descriptions of our main calculations.
Appendix~\ref{app:parameters} provides a discussion of the optimized protocol parameters.

\section{Conclusion}
We have devised a more advanced entanglement distribution protocol over thermal-loss channels, which improves upon a previously proposed qubit-bosonic strategy and integrates a recently developed entanglement distillation technique. 
Consequently, we have improved the lower bound for the two-way quantum capacity of the thermal-loss channel, a key open problem in quantum information theory.
Despite this improvement, the gap with the upper bound still remains large.
\par
It is unclear how much the lower bound can be further improved to close this gap. As we briefly discussed earlier, due to the computational complexity of the lookahead method of Ref.~\cite{HeuristicSearch}, we have not fully optimized the four parameters of our entanglement distribution protocol. Further improvements could therefore be obtained by optimizing these parameters fully.

In general, one potential approach to further improve the lower bound is to use a more powerful version of the lookahead protocol, for example by increasing the block size and search depth.
However, as Ref.~\cite{HeuristicSearch} shows, this would be computationally expensive.
In any case, an entanglement distillation procedure that, like ours, attempts to distribute a single entangled qubit pair per channel use can never achieve a rate greater than $1$.
Therefore, such a technique cannot be effective in the regime where the channel capacity is substantially greater than $1$. 
\par
Finally, similar efforts should also be directed to tightening the upper bound. Here, several strategies are possible, such as extending the minimization of the relative entropy of entanglement to include non-Gaussian states and/or adopting multi-copy regularizations.

\smallskip 
\noindent \textbf{Acknowledgements.}~We acknowledge support from UKRI via the Integrated Quantum Networks Research Hub (IQN, EP/Z533208/1). 

\FloatBarrier
\bibliography{references}

\appendix

\section{Distributed state}\label{OutputComputation}

\subsection{Thermal-loss action on a number-state dyad}\label{A0}

As a preliminary step, we need to describe the action of a thermal-loss channel on a number-state dyad, that is, on the outer product $\Ket{i}\Bra{j}$, where $\Ket{i}\Bra{i}$ is the $i$th number state.
First, let us introduce an auxiliary function $g$.
For all $x \in \left[0, 1\right]$, $y \in \left[1, \infty\right)$ and $\left(i, j, k, l\right) \in \left\{0, 1, 2, 3, \dots \right\}^{4}$ with $l < i$, $l < j$ and $j < k+l$, let $f_{i, j, k, l}\left(x, y\right)$ be 
\begin{equation}
     \frac{\sqrt{i!j!k!\left(i+k-j\right)!}\left(y-1\right)^{k+l-j}x^{\frac{i+j-2l}{2}}\left(1-x\right)^{l}}{\left(i-l\right)!\left(j-l\right)!l!\left(k+l-j\right)!y^{\frac{i+2k+2-j}{2}}}.
\end{equation}
Then, we define
\begin{equation}
    g_{i, j, k}\left(x, y\right) = \sum_{m = \max\left\{j-k, 0\right\}}^{\min\left\{i, j\right\}}{f_{i, j, k, m}\left(x, y\right)}\text{.}
\end{equation}
Now, given $\left(i, j\right) \in \left\{0, 1, 2, \dots\right\}^{2}$, one can write~\cite{LowerBound}
\begin{align}
     &\mathcal{E}_{\eta, \bar{n}}\left(\Ket{i}\Bra{j}\right)=\sum_{m=\max\left\{j-i, 0\right\}}^{\infty}g_{i, j, m}\left(x, y\right)\Ket{m+i-j}\Bra{m}\text{,} \nonumber \\
      &x = \frac{\eta}{1 + \left(1-\eta\right)\bar{n}},~~y = 1 + \left(1-\eta\right)\bar{n}\text{.}\label{ThermalLossAction}
\end{align}

\subsection{Computation of the distributed state}\label{A1}

Fix $m \in \left\{\max\left(\left\{0, N+1-M\right\}\right), \dots, N\right\}$.
Then using the action on a dyad, Eq.~\eqref{ThermalLossAction}, we can compute $p_{m}$ and $\rho^{\left(m\right)}$.
We have
\begin{equation}
    p_{m} = A_{m}+B_{m}+D_{m}+E_{m}
\end{equation}
and
\begin{equation}
    \rho^{\left(m\right)} = \frac{\sum_{\left(i, j\right) \in \left\{0, 1\right\}^{2}}{\Ket{e_{i}}\Bra{e_{j}}\sigma_{i, j}^{\left(m\right)}}}{p_{m}}\text{,}
\end{equation}
where
\begin{equation}
    A_{m} = c^{2}g_{M_{1}, M_{1}, m}\left(x, y\right)\text{,}
\end{equation}
\begin{equation}
    B_{m} = c^{2}g_{M_{1}, M_{1}, m+M}\left(x, y\right)\text{,}
\end{equation}
\begin{equation}
    C_{m} = c\sqrt{1-c^{2}}g_{M_{1}, M_{2}, m+M}\left(x, y\right)\text{,}
\end{equation}
\begin{equation}
    D_{m} = \left(1-c^{2}\right)g_{M_{2}, M_{2}, m}\left(x, y\right)\text{,}
\end{equation}
\begin{equation}
    E_{m} = \left(1-c^{2}\right)g_{M_{2}, M_{2}, m+M}\left(x, y\right)\text{,}
\end{equation}
\begin{equation}
    \sigma_{0, 0}^{\left(m\right)} = A_{m}\Ket{f_{0}}\Bra{f_{0}} + B_{m}\Ket{f_{1}}\Bra{f_{1}}\text{,}
\end{equation}
\begin{equation}
    \sigma_{0, 1}^{\left(m\right)} = C_{m}\Ket{f_{0}}\Bra{f_{1}}\text{,}
\end{equation}
\begin{equation}
    \sigma_{1, 0}^{\left(m\right)} = C_{m}\Ket{f_{1}}\Bra{f_{0}}
\end{equation}
and
\begin{equation}
    \sigma_{1, 1}^{\left(m\right)} = D_{m}\Ket{f_{0}}\Bra{f_{0}} + E_{m}\Ket{f_{1}}\Bra{f_{1}}\text{.}
\end{equation}

\subsection{Expansion in the Bell basis}\label{A2}

Let $\oplus$ represent addition modulo~$2$ and, for every $\left(i, j\right) \in \left\{0, 1\right\}^{2}$, let
\begin{equation}
    \Ket{\Phi_{i, j}} = \frac{\sqrt{2}\left(\Ket{e_{0}}\Ket{f_{i}}+\left(-1\right)^{j}\Ket{e_{1}}\Ket{f_{i\oplus1}}\right)}{2},
\end{equation}
so that $\left\{\Ket{\Phi_{0,0}}, \Ket{\Phi_{0,1}}, \Ket{\Phi_{1,0}}, \Ket{\Phi_{1,1}}\right\}$ is a Bell basis for a qubit pair.
For every $\left(i, j, k, l\right) \in \left\{0, 1\right\}^{4}$ and every $m \in \left\{\max\left(\left\{0, N+1-M\right\}\right), \dots, N\right\}$, we have
\begin{equation}
    \rho_{i, j, k, l}^{\left(m\right)} = \frac{F_{j, l, m}\delta_{i, 0}\delta_{k, 0}+G_{j, l, m}\delta_{i, 1}\delta_{k, 1}}{2p_{m}}\text{,}
\end{equation}
where $\delta$ is the Kronecker delta,
\begin{equation}
    F_{j, l, m} = A_{m} + \left(\left(-1\right)^{j}+\left(-1\right)^{l}\right)C_{m} + \left(-1\right)^{j+l}E_{m}
\end{equation}
and
\begin{equation}
    G_{j, l, m} = B_{m} + \left(-1\right)^{j+l}D_{m}\text{.}
\end{equation}
For every such $m$, the diagonal elements of $\rho^{\left(m\right)}$ in the Bell basis are therefore $\frac{A_{m}+2C_{m}+E_{m}}{2p_{m}}$, $\frac{A_{m}-2C_{m}+E_{m}}{2p_{m}}$, $\frac{B_{m}+D_{m}}{2p_{m}}$ and $\frac{B_{m}+D_{m}}{2p_{m}}$.

\begin{figure}[t!]
    \centering
    \includegraphics[width=0.91\columnwidth]{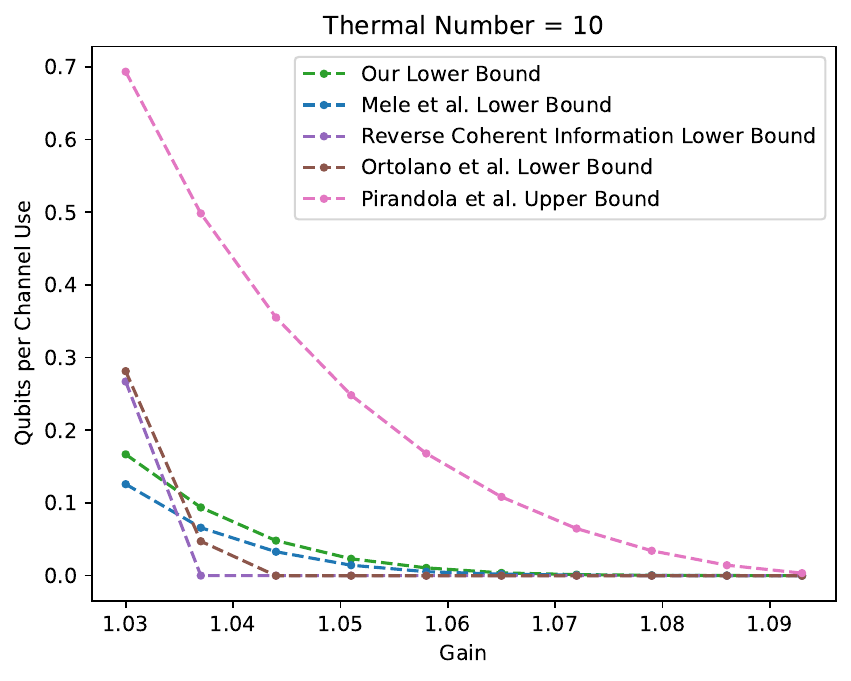}
    \caption{Bounds on the two-way quantum capacity of the quantum amplifier $\mathcal{E}_{g,\bar{n}}$. We plot the bounds versus the gain $g$ for $\bar{n}=10$. Our lower bound (green) is compared with the main lower bound from Ref.~\cite{LowerBound} (blue), the RCI lower bound of Ref.~\cite{ReverseCoherentInformation} (purple) and the PLOB upper bound for a quantum amplifier from Ref.~\cite{PLOB} (pink). For completeness, we also plot the secret-key capacity lower bound of Ref.~\cite{LB4} (brown). Dashed lines show linear interpolations between points; these are only intended as visual guides, not bounds themselves.}
    \label{fig:Tamplifier}
\end{figure}

\begin{figure}[t!]
    \centering
    \includegraphics[width=0.91\columnwidth]{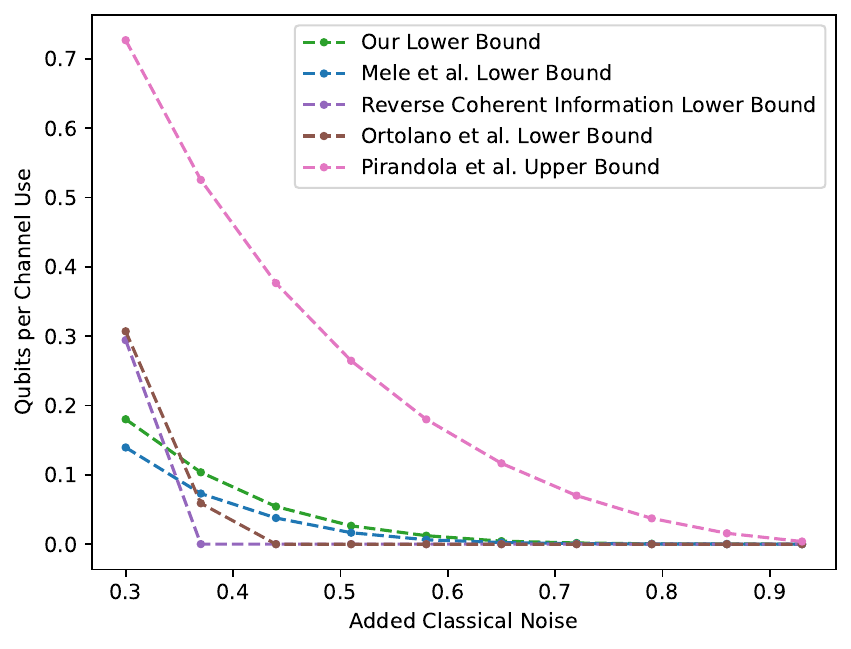}
    \caption{Bounds on the two-way quantum capacity of the Gaussian additive-noise channel $\mathcal{E}_{\xi}$. We plot the bounds versus the added noise variance $\xi$. Our lower bound (green) is compared with the main lower bound from Ref.~\cite{LowerBound} (blue), the RCI lower bound of Ref.~\cite{ReverseCoherentInformation} (purple) and the PLOB upper bound for the Gaussian additive noise from Ref.~\cite{PLOB} (pink). For completeness, we also plot the secret-key capacity lower bound of Ref.~\cite{LB4} (brown). Dotted lines show linear interpolations between points; these are only intended as visual guides, not bounds themselves.}
    \label{fig:Additive}
\end{figure}

\section{Results for the quantum amplifier and the Gaussian additive-noise channel}\label{app:gchannel}

Here we show that our results can be extended to other bosonic Gaussian channels.
The first channel we consider is the quantum amplifier $\mathcal{E}_{g,\bar{n}}$, whose action on an input quadrature $\hat{x}$ is given by $\hat{x} \rightarrow \sqrt{g} \hat{x} + \sqrt{g-1} \hat{x}_E $, where $g>1$ is the gain and $\hat{x}_E$ is an environmental mode in a thermal state with $\bar{n}$ mean photons.
The second channel is the Gaussian additive-noise channel $\mathcal{E}_{\xi}$, whose action is $\hat{x} \rightarrow \hat{x} + z$, where $z$ is a random Gaussian variable with zero mean and variance $\xi >0$.
\par
Our lower bounds for the two-way quantum capacity of these channels can be calculated in much the same way as in the thermal-loss case.
As shown in Ref.~\cite{LowerBound}, the only required change is to the values of $x$ and $y$, defined in Eq.~\eqref{ThermalLossAction}.
In the case of the quantum amplifier $\mathcal{E}_{g,\bar{n}}$, we must replace $x$ and $y$ with
\begin{equation}
    x_{\text{amplifier}} = \frac{g}{g + \left(g-1\right)\bar{n}}, ~~y_{\text{amplifier}} = g + \left(g-1\right)\bar{n}.
\end{equation}
In the case of the Gaussian additive-noise channel $\mathcal{E}_{\xi}$, we replace $x$ and $y$ with
\begin{equation}
    x_{\text{additive}} = \frac{1}{1+\xi},~~y_{\text{additive}} = 1 + \xi.
\end{equation}
Then, the calculations follow the same steps as in Appendices~\ref{A1} and~\ref{A2}. 
\par
As shown in Figs.~\ref{fig:Tamplifier} and~\ref{fig:Additive}, our improved lower bound outperforms the result of Ref.~\cite{LowerBound}, reducing the gap between lower and upper bounds for the two-way quantum capacity $Q_{2}$ of these important Gaussian channels.
Furthermore, we improve upon the results of Ref.~\cite{LB4}, the best known lower bound on $K_{2}$, at higher values of the gain or added noise.

\begin{figure}[b!]
\centering
\includegraphics[width=0.48\textwidth]{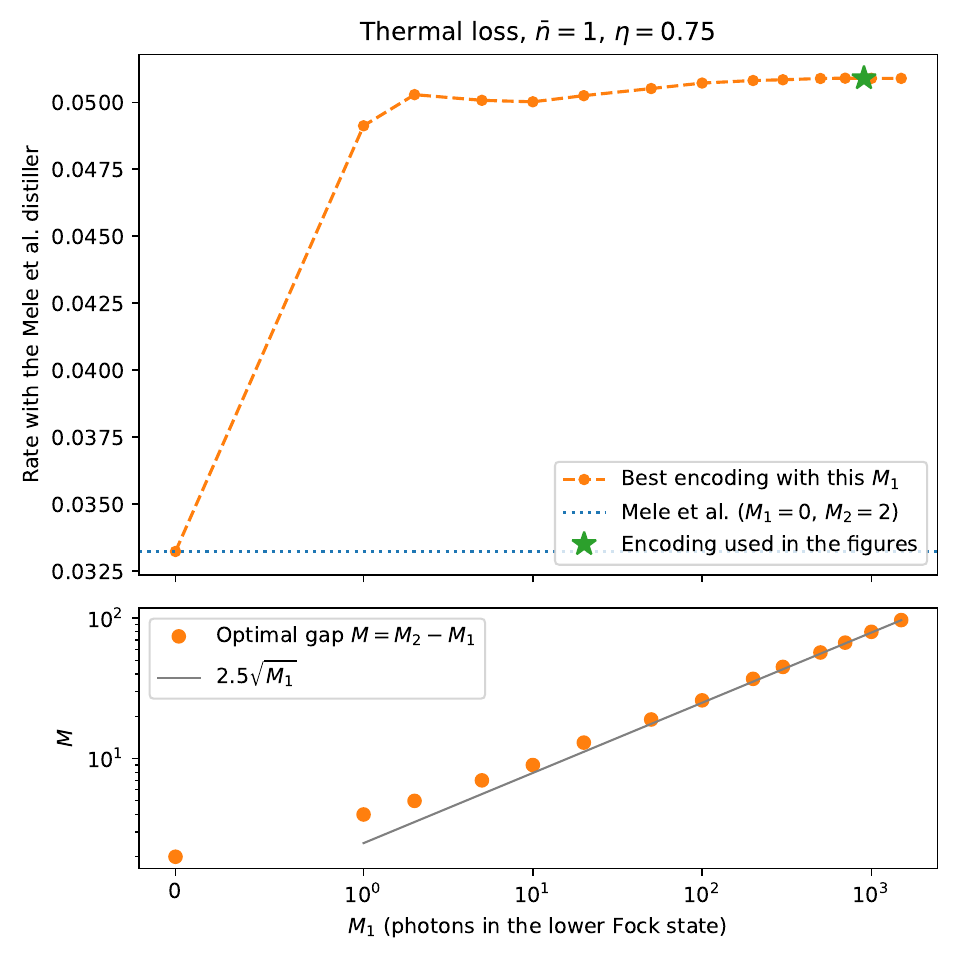}
\caption{Thermal-loss channel with $\bar n=1$ and $\eta=0.75$. Top: best rate with the
distiller of Ref.~\cite{LowerBound} for fixed $M_1$, optimized over $c$, $M_2$ and $N$
(orange); the encoding of Ref.~\cite{LowerBound} (blue, dotted); the encoding used in
Fig.~1 (green star). Bottom: optimal gap $M=M_2-M_1$.}
\label{fig:degeneracy}
\end{figure}

\begin{algorithm*}[!t]
\caption{Yield of the distiller of Ref.~\cite{LowerBound}, $Y_{\rm M}(\vec q)$}
\label{alg:mele}
\begin{algorithmic}[1]
\Require Bell-diagonal weights $\vec q$, written as $\alpha_{ab}$ ($a$ phase, $b$ bit flip)
\State $J(\alpha)\gets$ maximum two-copy yield of Ref.~\cite{PhysRevA.71.062325} (Eq.~(S21) of Ref.~\cite{LowerBound}) over the Bell relabellings of Ref.~\cite{LowerBound}
\State $Y\gets J(\alpha)$,\quad $f\gets1$
\For{$t=1,\dots,25$}
  \State apply one P1-or-P2 round (Eqs.~(S126)--(S129) of Ref.~\cite{LowerBound}): P1 if
        $\alpha_{10}<\alpha_{01}$, otherwise P2; success probability $P_t$
  \State $f\gets f\,P_t/2$; \quad \textbf{if} $f<10^{-12}$ \textbf{break}
  \State $Y\gets\max\big(Y,\ f\,J(\alpha)\big)$
\EndFor
\State \Return $Y$
\end{algorithmic}
\end{algorithm*}

\begin{algorithm*}[!t]
\caption{Rate of the encoding $\theta=(c,M_1,M_2,N)$}
\label{alg:rate}
\begin{algorithmic}[1]
\Function{MeleRate}{$\theta$} \Comment{orange curve; objective of the optimization}
  \State \Return $\sum_m p_m\,Y_{\rm M}(\vec q^{(m)})$ over classes with $p_m\ge10^{-9}$
\EndFunction
\Function{Rate}{$\theta$} \Comment{green curve}
  \For{each class $m$}
    \State $y_m\gets Y_{\rm M}(\vec q^{(m)})$
    \State $\vec s_0\gets\vec q^{(m)}$, $f_0\gets1$, $b\gets0$
    \For{$k=0,1,\dots,10$}
      \If{$\max\vec s_k\le\tfrac12$ \textbf{or} $f_k\,[1-h_2(\max\vec s_k)]\le b$}
        \textbf{break} \EndIf
      \State $b\gets\max\big(b,\ f_k\,Y_{\rm LA}(\vec s_k)\big)$
             \Comment{lookahead, $(n,r,d)=(4,2,3)$}
      \State $(\vec s_{k+1},N_k)\gets\textsc{Dejmps}(\vec s_k)$;\quad $f_{k+1}\gets f_kN_k/2$
    \EndFor
    \State $y_m\gets\max(y_m,b)$
  \EndFor
  \State \Return $\sum_m p_m\,y_m$
\EndFunction
\end{algorithmic}
\end{algorithm*}

\begin{table}
\vspace{-0.2cm}
\caption{Optimized parameters for the thermal-loss channel with $\bar n=1$. $R_{\rm M}$:
rate with the distiller of Ref.~\cite{LowerBound} (blue curve for the encoding of
Ref.~\cite{LowerBound}; orange curve for ours). $R$: our lower bound (green curve).}
\label{tab:params}
\centering
\begin{tabular}{c|ccc|cccccc}
\hline
 & \multicolumn{3}{c|}{Ref.~\cite{LowerBound}} & \multicolumn{6}{c}{This work} \\
$\eta$ & $c$ & $M$ & $R_{\rm M}$ & $c$ & $M_1$ & $M_2$ & $N$ & $R_{\rm M}$ & $R$ \\
\hline
0.55 & 0.784 & 1 & 0.00001 & 0.643 & 2 & 6 & 1 & 0.00001 & 0.00007 \\
0.57 & 0.767 & 1 & 0.00004 & 0.698 & 1 & 4 & 0 & 0.00007 & 0.00025 \\
0.59 & 0.774 & 1 & 0.00013 & 0.705 & 1 & 4 & 1 & 0.00030 & 0.00074 \\
0.61 & 0.794 & 2 & 0.00040 & 0.676 & 1 & 4 & 1 & 0.00077 & 0.00169 \\
0.63 & 0.787 & 2 & 0.00091 & 0.670 & 1 & 4 & 1 & 0.00161 & 0.00318 \\
0.65 & 0.753 & 2 & 0.00201 & 0.681 & 1 & 4 & 1 & 0.00353 & 0.00627 \\
0.67 & 0.774 & 2 & 0.00422 & 0.706 & 481 & 540 & 334 & 0.00685 & 0.01134 \\
0.69 & 0.769 & 2 & 0.00771 & 0.706 & 1301 & 1398 & 917 & 0.01233 & 0.01868 \\
0.71 & 0.725 & 2 & 0.01273 & 0.676 & 1 & 5 & 1 & 0.02054 & 0.03001 \\
0.73 & 0.713 & 2 & 0.02221 & 0.676 & 1 & 5 & 1 & 0.03363 & 0.04469 \\
0.75 & 0.703 & 2 & 0.03323 & 0.706 & 904 & 980 & 713 & 0.05090 & 0.06582 \\
0.77 & 0.684 & 3 & 0.05342 & 0.706 & 680 & 745 & 546 & 0.07369 & 0.09355 \\
\hline
\end{tabular}
\end{table}

\section{Numerical methods}
\label{app:numerics}

Algorithms \ref{alg:mele}, \ref{alg:rate} and \ref{alg:opt} describe how we calculated our lower bounds.
Algorithm~\ref{alg:mele} describes Ref.~\cite{LowerBound}'s distillation protocol, Algorithm~\ref{alg:rate} describes how we calculated the yields of our protocols once we had chosen $\left(c, M_{1}, M_{2}, N\right)$ and Algorithm~\ref{alg:opt} describes how we optimized these parameters.
There is some randomness in our choice of where to begin our local parameter optimizations from and there is no guarantee that Algorithm~\ref{alg:opt} will find the globally optimal parameters.
Throughout, \textsc{DEJMPS} refers to the purification protocol of Ref.~\cite{Recurrence}.

\begin{algorithm*}
\caption{Optimization of $(c,M_1,M_2,N)$}
\label{alg:opt}
\begin{algorithmic}[1]
\Function{F}{$M_1,M_2,N$} \Comment{$c$ optimized for each integer point}
  \State maximize \textsc{MeleRate} over $c\in(0,1)$ (grid scan then bounded Brent refinement)
  \EndFunction
\Function{Descend}{$x$} \Comment{$x=(M_1,M_2,N)$, feasible iff $0\le M_1<M_2$, $N\ge0$}
  \Loop
    \State $x'\gets x$,\quad $\varepsilon\gets\max(10^{-15},\,10^{-6}|F(x)|)$
    \For{$e$ in $(1,0,0),(0,1,0),(0,0,1),(1,1,0),(0,1,1),(1,1,1)$}
      \For{$s=1,2,4,\dots,64$, and for each $s$ first $+s$ then $-s$}
        \If{$x+s\,e$ feasible \textbf{and} $F(x+s\,e)>F(x')+\varepsilon$}
          $x'\gets x+s\,e$ \EndIf
      \EndFor
    \EndFor
    \If{$x'=x$} \Return $x$ \EndIf
    \State $x\gets x'$
  \EndLoop
\EndFunction
\Function{Restarts}{$x_0$} \Comment{12 restarts}
  \State $x^\star\gets\textsc{Descend}(x_0)$;\quad $S\gets10$
  \For{$t=1,\dots,12$}
    \State $x_i\gets\operatorname{round}(x^\star_i S^{u_i})$ with $u_i$ uniform on $[-1,1]$ (if $x^\star_i=0$: $x_i$ uniform on $\{0,\dots,\lceil S\rceil\}$)
    \State sort $M_1,M_2$;\ $M_2\gets\max(M_2,M_1+1)$    
    \State $x\gets\textsc{Descend}(x)$
    \If{$F(x)>F(x^\star)\,(1+10^{-3})$} $S\gets\min(2S,1000)$
    \Else\ $S\gets\max(S/2^{1/4},1.5)$ \EndIf
    \State \textbf{if} $F(x)>F(x^\star)$ \textbf{then} $x^\star\gets x$  \EndFor
  \State \Return $x^\star$
\EndFunction
\State run \textsc{Restarts} from several points, including $(0,M,0)$ with $M$ the optimum over $M=1,\dots,40$, and $(1,5,1)$
\State \Return the best result
\end{algorithmic}
\end{algorithm*}

\section{Optimized parameters}
\label{app:parameters}

Table~\ref{tab:params} lists, for the thermal-loss channel with $\bar n=1$, the parameters
behind every plotted point: the optimal encoding $(c,M)$ of Ref.~\cite{LowerBound}, with
$M_1=0$ and $M_2=M$, and our optimal encoding $(c,M_1,M_2,N)$, which is shared by the orange
and green curves.
\par
Although our optimized values sometimes involved much larger photon numbers than those of Ref.~\cite{LowerBound}'s protocol, similar improvements can be obtained with small photon numbers.
Figure~\ref{fig:degeneracy} shows, at $\eta=0.75$, the best rate with the distiller of Ref.~\cite{LowerBound} when $M_1$ is held fixed and the remaining parameters are optimized.
Moving from $M_1=0$, the encoding of Ref.~\cite{LowerBound}, to $M_1=2$ raises the rate by about $50\%$; beyond that the rate changes by only about $1\%$ up to $M_1=1500$, with the optimal gap $M=M_2-M_1$ growing approximately as $2.5\sqrt{M_1}$ and the cutoff remaining at about $0.77\,M_1$.
Because the rate is so flat in the photon number, the optimizer settles on small Fock states at some transmissivities and on Fock states with hundreds of photons at others, as Table~\ref{tab:params} shows.
These choices are interchangeable to within about $1\%$ in rate; in particular, encodings with $M_1$ of order one achieve almost the entire improvement over Ref.~\cite{LowerBound}.

\end{document}